\documentclass[twocolumn,showpacs,pre,aps]{revtex4}

\usepackage{graphicx}
\usepackage{dcolumn}
\usepackage{bm}

\begin{document}

\title{On Statistics and 1/f Noise of Brownian Motion\\
in Boltzmann-Grad Gas and Finite Gas on Torus. II. Finite Gas}

\author{Yuriy E. Kuzovlev}
\email{kuzovlev@kinetic.ac.donetsk.ua} \affiliation{Donetsk Institute
for Physics and Technology of NASU, 83114 Donetsk, Ukraine}


\begin{abstract}
An attempt is made to compare statistical properties of
self-diffusion of particles constituting gases in infinite volume and
on torus. In this second part, derivation, from BBGKY equations, of
roughened model of self-diffusion is revised as applied to finite
$N$-particle gas under micro-canonical ensemble. The model confirms
existence of characteristic time $\approx N$, in units of free flight
time, for cross-over between non-Gaussian and Gaussian regimes of
diffusion, but then loses its legacy.
\end{abstract}


\maketitle

\section{Introduction}

In the first part \cite{p1} of present work a most simplified version
of the ``collisional approximation'' \cite{i1} of the
Bogolyubov-Born-Green-Kirkwood-Yvon (BBGKY) equations was formulated
and then used to analyze statistics of Brownian motion of gas
particles in an infinite-volume gas under the Boltzmann-Grad limit
($\,\nu\rightarrow\infty\,$, $\,\delta\rightarrow 0\,$,
$\,\lambda\sim 1/\nu\delta^{\mathcal{D}-1} =\,$const\,, where
$\,\mathcal{D}\,$ is space dimension, $\,\nu\,$ and $\,\delta\,$ are
concentration of gas particles and radius of their repulsive
interaction, respectively, and $\,\lambda\,$ their mean free path).
Now, our aim is to do something similar in respect to finite-volume
gas on torus with finite total number of particles $\,N\,$.

Recall that the ``collisional approximation'' was introduced in
\cite{i1} (one can see also \cite{i2} or Appendix B in \cite{p1})
being mentioned as correct transition from exact BBGKY theory to
Boltzmann-like theory in the case of spatially nonuniform gas, and in
such sense it pretends to realization of earlier outspoken doubts
about validity of the Boltzmann equation \cite{kac} or, equivalently,
provability of Boltzmann's ``Stosshalansatz'' \cite{rdl} in this
case. One of key observations in \cite{i1} was that relative motion
of colliding particles is inner constituent of their collision and
therefore should be excluded from its outer characterization. In
other words, probability density (ensemble-average concentration) of
pair collisions $\,F_2\,$ undergoes an equation $\,\partial
F_2/\partial t =-\bm{V}\cdot\partial F_2/\partial \bm{R}+...\,$,
where $\,\bm{V}=(\bm{v}_1+\bm{v}_2)/2\,$ and
$\,\bm{R}=(\bm{r}_1+\bm{r}_2)/2\,$ are velocity and position of the
collision (more precisely, its center of gravity), and the dots
replace ``collision operators'' acting onto the velocities. It is
easy to prove that in actually nonuniform situation those $\,F_2\,$
(in contrary to general-position pair distribution function for
non-colliding particles) can not be factored into product
$\,F_1(t,\bm{r}_1\approx\bm{R},\bm{v}_1)$$
F_1(t,\bm{r}_2\approx\bm{R},\bm{v}_2)\,$ of two one-particle
distribution functions. Hence, it turns out to be independent
supplementary characteristics of non-uniform gas.

If that is the case, then concentration $\,F_3\,$ of three-particle
clusters, or ``encounters'' \cite{i1}, which enters the dots above,
also is independent. It in its turn involves four-particle
encounters, and so on. Thus we have to consider many coupled linear
kinetic equations, which could be reduced to the single nonlinear
Boltzmann equation under strict spatial uniformity only.

As in \cite{i1} (or \cite{i2} or \cite{p1}) we will deal with not
``thermodynamical'' but ``statistical'' (``informational'') spatial
non-uniformity implied by information about that (at some initial
time moment $\,t=0\,$) one of gas particles is definitely positioned
near some given place $\bm{r}=\bm{r}_0\,$. Corresponding (normalized
to unit) distribution function of the whole gas can be chosen as
\begin{eqnarray}
F_N(\bm{r},\bm{v})\,= \,F_N^{\,eq}(\bm{r},\bm{v})\,\frac
{\Omega}{N}\,\sum_{j=1}^N
\,\delta^{(\mathcal{D})}(\bm{r}_j-\bm{r}_0)\,\,\,,
\,\,\,\,\,\,\,\label{ind}\\
F_N^{\,eq}(\bm{r},\bm{v})\,=\, \frac
{C_N(E)}{\Omega^{\,N}}\,\,\delta\left(E_N(\bm{r},\bm{v})\,-E\right)
\delta^{(\mathcal{D})}\left(\sum_{j=1}^N\bm{v}_j\right) \,,\nonumber
\end{eqnarray}
where $\,\Omega=l^{\mathcal{D}}\,$ is volume of gas contained in
``flat cubic'' torus $\,0\leq r_{\alpha}\leq l\,$ ($\,\alpha=1\div
\mathcal{D}\,$), $\,E_N(\bm{r},\bm{v})\,$ is total energy of gas, and
$\,F_N^{\,eq}(\bm{r},\bm{v})\,$ its equilibrium distribution with
$\,C_N(E)\,$ being normalizing multiplier. Thus, unlike \cite{p1}, we
will rest upon micro-canonical ensemble, with fixed total energy and
zero total momentum, which looks natural at finite number of degrees
of freedom.

We want to study consequent evolution of $\,F_N\,$ in order to obtain
statistical characteristics of Brownian motion of initially localized
particle. Unfortunately, direct solution of exact Liouville equation
for $\,F_N\,$ is impossible. From the other hand, the Boltzmann
equation (or, rather, Boltzmann-Lorentz equation \cite{rdl}) for
$\,F_1\,$ can be easy solved, but in view of above reasonings (see
\cite{i1} for detail discussion) it seems inadequate with respect to
spatial dependence of $\,F_1\,$. Therefore by perforce we have to use
the theory from \cite{p1} adapting it to finite system. At that, we
can not exploit the Boltzmann-Grad limit which helps to obviate most
of difficulties of the ``collisional approximation'' descending from
vagueness of the very concept of ``collision'' \cite{i1}. By these
reasons, at present only quite imperfect and caricature collisional
model of nonuniform finite gas is available. Nevertheless, may be, it
is better than uncritical use of the Boltzmann equation, all the
more, better than nothing.

\section{Model equations}

{\bf 1.\,} Let us make non-principal assumptions as follow. First, of
course, that total number of particles is large enough, $\,N\gg 1\,$.
Second, mean free path is comparable with the torus dimensions:
$\,\lambda\sim 1/\nu\delta^{\mathcal{D}-1} \sim l\,$, with $\,\nu
=N/\Omega =N/l^{\mathcal{D}}\,$. Under these assumptions, estimate of
a number of simultaneously occurring collisions obviously yields a
value much smaller than unit:
$\,N(\delta^{\mathcal{D}}N/\Omega)/2\,\sim\,
N^{-1/(\mathcal{D}-1)}\,\ll 1\,$. Hence, our gas is sufficiently
rarefied. On this ground, when considering probabilities of particle
velocities, we can neglect potential energy contribution to
$\,E_N(\bm{r},\bm{v})\,$ and write in (\ref{ind})
$\,E=N\mathcal{D}Mv_0^2/2\,$, with $\,v_0\,$ being characteristic
velocity ($\,M\,$ is particle mass).

In order to get ability of watching for full rotations of particles
around torus, let us use extended configurational space $\,-\infty
<r_{\alpha}<\infty\,$, so that its cubic cell $\,(m_{\alpha}-1/2)l
<r_{\alpha}<(m_{\alpha}+1/2)l\,$, with integer numbers
$\,m_{\alpha}$, corresponds to $\,m_{\alpha}\,$ rotations along
$\,\alpha$-th dimension during time period after $\,t=0\,$. Under
this representation of life on torus, potential energy of interaction
of any two particles changes to periodic function of coordinates,
$\,U(\bm{\rho}(\bm{r}_i-\bm{r}_j))\,$, where $\,\bm{\rho}(\bm{r})\,$
means $\,\bm{r}\,$ modulo $\,l\,$:
$\,\bm{\rho}_{\alpha}(\bm{r})\,$$=\bm{r}_{\alpha}-m_{\alpha}l\,$ when
$\,(m_{\alpha}-1/2)l <r_{\alpha}<(m_{\alpha}+1/2)l\,$. At that,
distribution (\ref{ind}) also will be extended to the whole space
$\,-\infty <r_{\alpha}<\infty\,$, with $\,\Omega\,$ now being
mentioned as its (infinite) volume. Such statistical ensemble
implies, clearly, that rotations of only a sole (initially localized)
particle will be watched, while getting of information about absolute
positions and displacements of other particles is beforehand
eliminated.

Next, consider particular $\,n$-particle distribution functions which
follow from (\ref{ind}):
\begin{eqnarray}
F_n(\bm{r}_1...\,\bm{v}_n)\,= \,\int_{n+1}...\int_N
F_N(\bm{r},\bm{v})\,=\,\,\,\,\,\,\,\,\,\,\,\,\,\,\nonumber\\
=\,\frac {\Omega}{N}\,\,F_n^{\,eq}(\bm{r}_1...\,\bm{v}_n)
\sum_{j=1}^n
\,\delta^{(\mathcal{D})}(\bm{r}_j-\bm{r}_0)\,+\,\,\,\,\,\,\,\label{fni}\\
\,\,\,\,\,+\,\frac {(N-n)\,\Omega}{N}\,\int
F_{n+1}^{\,eq}(\bm{r}_1...\,\bm{r}_n,\bm{r}_0,\bm{v}_1...\,\bm{v}_n,
\bm{v}^{\prime})\,d\bm{v}^{\prime}\,\,\,,\nonumber\\
 F_n^{\,eq}(\bm{r}_1...\,\bm{v}_n)\,=\,
\int_{n+1}...\int_N\,F_N^{\,eq}(\bm{r},\bm{v})
\,\,\,,\,\,\,\,\,\,\,\,\,\,\,\label{eqf}
\end{eqnarray}
where $\,\int_s...=\int...\,\,d\bm{v}_sd\bm{r}_s\,$. Since, according
to (\ref{ind}), $\,F_n\sim \Omega^{-\,n}\,$, and $\,\Omega\,$ has
become infinite, while $\,N\,$ is finite, it is evident that the
second term on right-hand side of (\ref{fni}) factually vanishes in
comparison with first term and should be thrown away when considering
nonuniform (localized) parts of $\,F_n$'s. Besides, it is reasonable
to move off also common multiplier $\,1/N\,$. Thus we can write
\begin{eqnarray}
F_n(\bm{r}_1...\,\bm{v}_n)\,= \,
\Omega\,F_n^{\,eq}(\bm{r}_1...\,\bm{v}_n)\, \sum_{j=1}^n
\,\delta^{(\mathcal{D})}(\bm{r}_j-\bm{r}_0)\,\label{fni1}
\end{eqnarray}
with normalization $\,\int_1...\int_n\,F_n\,=\,n\,\Omega^{\,n-1}\,$.

{\bf 2.\,} Formula (\ref{fni1}), along with (\ref{ind}) and
(\ref{eqf}), presents initial conditions for the BBGKY equations
\begin{eqnarray}
\frac {\partial F_n}{\partial t}\,=\,L^{(n)}F_n+(N-n)\int_{n+1}
\sum_{j=1}^nL_{j\,n+1}F_{n+1}\,\,\,,\label{bfn}\\
L^{(n)}\,=\,-\sum_{j=1}^n \bm{v}_j\cdot\frac {\partial }{\partial
\bm{r}_j}\,+\sum_{1\leq\, i<j\,\leq n} L_{\,ij}
\,\,\,,\,\,\,\,\,\,\,\,\,\nonumber\\
\,\,\,\,\,\,L_{\,ij}\,=\, \nabla
U(\bm{\rho}(\bm{r}_i-\bm{r}_j))\cdot\left(\frac {\partial }{\partial
\bm{p}_{\,j}}-\frac {\partial }{\partial
p_i}\right)\,\,\,\,\,\,\,\,\,\,\,\,\,\,\nonumber
\end{eqnarray}
In ``collisional approximation'' we want to transform the right-hand
integrals in (\ref{bfn}) into Boltzmann-like ``collisional
integrals''.

This is possible in those cases only when statistical ensemble under
consideration contains all various consecutive phases (instant
dynamic states) of a collision process, moreover, when all that
phases are represented by the ensemble with equal probability.
Otherwise, the collision process can not be replaced by momentary
collision event, since the latter must conserve probability (in
essence, conserve particles). Of course, hardly a non-trivial
ensemble ever (even at Bogolyubov's ``kinetic stage'') undergoes such
requirement in formally rigorous sense. Therefore this requirement
should be interpreted as instruction how to make ``collisional
approximation''.

Firstly, we must consider special values of $\,F_n\,$ ($\,n>1\,$)
corresponding to clusters (``encounters'' \cite{i1,i2}) of $\,n\,$
mutually close particles (e.g. satisfying, in real torus space,
conditions $\,|\bm{r}_i-\bm{r}_j|<d\,$, where $\,d\,$ is
significantly smaller than $\,l\,$ but greater than $\,\delta\,$). At
$\,n=2\,$ the matter concerns either literally collision or close
encounter of two particles with no essential interaction, while at
$\,n>2\,$ we speak about chains of close pair collisions or their
combinations with mere encounters (in particular, might-have-been or
``virtual'' collisions in the sense mentioned in \cite{kac} or
\cite{rdl}).

Secondly, divide term $\,L^{(n)}F_n\,$ in (\ref{bfn}) into two parts,
one coming from non-uniformity of statistical ensemble, that is
$\,F_n$'s dependence on translations $\,\bm{r}_j\rightarrow\bm{r}_j
+\Delta\bm{r}\,$, and other from its dependence on dynamic phase of
$\,n$-particle encounter. This can be done in the form
\begin{eqnarray}
L^{(n)}F_n\,=-\bm{V}^{(n)}\cdot\sum_{j=1}^n \frac {\partial
F_n}{\partial \bm{r}_j}-\frac {\partial F_n}{\partial \Theta}\,\,\,,
\label{ln}
\end{eqnarray}
where $\,\bm{V}^{(n)}=(\bm{v}_1+...+\bm{v}_n)/n\,$ is velocity of
centroid of the $\,n$-particle cluster, and $\,\Theta\,$ is its inner
time in the centroid's reference frame. Obviously, $\,\sum
\partial /\partial \bm{r}_j\,$ is just infinitesimal
operator of plane-parallel displacement of the whole cluster. About
$\,\Theta\,$ see \cite{i2} or Appendix B in \cite{p1}.

Thirdly, and chiefly, following the above requirement, second term in
(\ref{ln}) must pass for zero: $\,\partial F_n/\partial
\Theta\rightarrow 0\,$, since the concept of collisions imposes
equiprobability of its dynamic phases (in particular, its border
``in-'' and ``out-'' states), i.e. conservation of probability.

After all, $\,F_n\,$, being taken at
$\,|\bm{\rho}(\bm{r}_i-\bm{r}_j)|<d\,$, becomes probabilistic
characteristics of $\,n$-particle encounters as one-piece events. Two
right-hand terms in (\ref{bfn}) turn into drift and source of the
encounters, respectively (to be precise, drift and source of their
ensemble-average concentration). At that, the requirement $\,\partial
F_n/\partial \Theta\rightarrow 0\,$ once again plays important role:
it serves for transforming the source into sum of collision operators
which act upon $\,F_{n+1}\,$ \cite{i1,i2}.

However, still $\,F_n\,$ possesses too many arguments (after removal
of $\,\Theta\,$, only by unit less than generally), including impact
parameters and other internal geometric details of an encounter.
Therefore simultaneously let us exclude the latter arguments by
averaging $\,F_n\,$ over them. Beforehand notice that spatial
dependence of exact solution to (\ref{bfn}) starting from
(\ref{fni1}) must look like
\begin{eqnarray}
F_n\,=\,\widetilde{F}_n(t,\bm{r}_1\,;\,
\bm{\rho}(\bm{r}_2-\bm{r}_1),...\,
\bm{\rho}(\bm{r}_n-\bm{r}_1))+\,\text{CP}\,\,\,,\label{es}
\end{eqnarray}
where $\,\widetilde{F}_n\,$ is symmetric with respect to indices
$\,2\div n\,$ and ``CP'' means sum of $\,n-1\,$ cyclic permutations
of all $\,n\,$ indices. For this reason, it is sufficient (with no
loss of information) to average $\,F_n\,$ over only $\,\bm{r}_j\,$
belonging to one and the same cell of the extended space (and
multiply results by some constants to keep suitable normalization).

Thus, fourthly, introduce $\,W(\bm{r})\,$ as some spherically
symmetric bell-shaped (and normalized to unit) weighting function
with width $\,\sim d\,$ (i.e. approximately enclosing a region
assigned to collisions and encounters) and then averaged
distributions as
\begin{eqnarray}
\overline{F}_n\,=\,\Omega^{n-1}\int...\int F_n\,\prod_{j=1}^n
W(\bm{r}_j-\bm{R})\,d\bm{r}_j\,\,\, \label{afn}
\end{eqnarray}
Applying such operation to (\ref{bfn}), in view of (\ref{es}), we can
replace expressions $\,\int
d\bm{r}_{n+1}\,U(\bm{\rho}(\bm{r}_j-\bm{r}_{n+1}))\,...\,$ by
expressions $\,(\nu\,\Omega/N)\,\int
d\bm{r}_{n+1}\,U(\bm{r}_j-\bm{r}_{n+1})\,...\,$ with $\,\Omega\,$
being (infinite) volume of the extended space and $\,\nu\,$ (real
finite) concentration of gas. Taking into account also equalities
(\ref{ln}) and $\,\partial F_n/\partial \Theta =0\,$, and
manipulating by analogy with \cite{i1,i2}, it is not hard to obtain
\begin{eqnarray}
\frac {\partial \overline{F}_n}{\partial t}=\,-\bm{V}^{(n)}\cdot\frac
{\partial \overline{F}_n}{\partial \bm{R}}+\frac
{N-n}{N}\,\sum_{j=1}^n
\,S_{j\,n+1}\overline{F}_{n+1}^{\,in}\,\label{aeq}
\end{eqnarray}
with $\bm{V}^{(n)}=n^{-1}\sum_{j\,=1}^n\bm{v}_j\,$\,. Symbol
$\,S_{j\,n+1}\,$ designates Boltzmannian linear collision operator
describing collision between $\,j$-th member of $\,n$-particle
cluster and an exterior $\,(n+1)$-th particle. The latter came from
outside of the cluster and is in incoming state with respect to it,
being placed somewhere at its border. Function
$\,\overline{F}_{n+1}^{\,in}\,$ just represents such situation.

The last collision terms in (\ref{aeq}) split to purely pair
collisions (without participation of other members of a cluster at
$\,n>2\,$). Strictly speaking, this statement is simply demonstrable
only for infinite gas under the Boltzmann-Grad limit (when $\,d\,$
can be chosen so that $\,\delta/d\rightarrow 0\,$ despite
$\,d/\lambda\rightarrow 0\,$ and any cluster becomes ``infinitely
rarefied from within'' although``infinitely dense from without'').
Nevertheless, even if it is formally wrong, in essence this is not
significant defect of our model, because, as we will see, factual
role of collisions is exchange between momentum of ``outer
$(n+1)$-th'' particle and total momentum of $\,n$-particle cluster.

What is more important, we came to the point where, fifthly, we need
in a hypothesis like Boltzmann's Stosshalansatz (``molecular chaos'')
in order to reduce functions $\,\overline{F}^{\,in}\,$ to functions
$\,\overline{F}$ and thus close equations (\ref{aeq}). But both
trivial reasoning mentioned in the Introduction and deeper
considerations \cite{i1,i2} show that it can not be literally
Boltzmann's hypothesis. If speak about infinite gas, nothing prevents
a cluster's (``$j$-th'') particle and exterior (``$(n+1)$-th'')
particle be statistically independent in the sense of their
velocities, but, naturally, their positions are statistically
dependent, which is expressed by
$\,\overline{F}_{n+1}^{\,in}(t,\bm{R},\bm{v}_1...\,\bm{v}_n,\bm{v}_{n+1})$
$=\,F_0(\bm{v}_{n+1})$$\int
\overline{F}_{n+1}(t,\bm{R},\bm{v}_1...\,\bm{v}_n,
\bm{v}^{\prime})\,d\bm{v}^{\prime}\,$ (with $\,F_0(\bm{v})\,$ being
Maxwell distribution when gas is in thermodynamical equilibrium).
However, in finite gas, framed by the micro-canonical ensemble
(\ref{ind}), even velocities of ``$j$-th'' and ``$(n+1)$-th''
particle must be statistically dependent. In view of this new
difficulty, relations of $\,\overline{F}_{n+1}^{\,in}\,$ to
$\,\overline{F}_{n+1}\,$ will be analyzed in concrete way of further
simplification of the model.

{\bf 3.\,} According to (\ref{ind}), (\ref{fni1}) and (\ref{afn}) and
notions made in the beginning of this Section, initial conditions to
equations (\ref{aeq}) must be
\begin{eqnarray}
\overline{F}_n(t=0,\bm{R},\bm{v}_1...\,\bm{v}_n)\,=
\,n\,W(\bm{R})\,\overline{F}_n^{\,eq}(\bm{v}_1...\,\bm{v}_n)
\,\,\,,\label{aind}\\
\overline{F}_n^{\,eq}(\bm{v}_1...\,\bm{v}_n)\,=\,\int
d\bm{v}_{n+1}\,...\int d\bm{v}_N
\,\times \,\,\,\,\,\,\,\,\,\,\,\,\,\,\,\,\nonumber\\
\times\,\,C_N(E)\,\delta\left(\frac M2
\sum_{j\,=1}^N\bm{v}_j^2\,-E\right)\delta^{(\mathcal{D})}
\left(\sum_{j\,=1}^N\bm{v}_j\right)\,\,\,,\nonumber
\end{eqnarray}
where $\,E=N\mathcal{D}Mv_0^2/2\,$, we took $\,\bm{r}_0=0\,$, and
$\,\overline{F}_n^{\,eq}\,$ are equilibrium velocity distributions
normalized to unit. The integration yields
\begin{eqnarray}
\overline{F}_n^{\,eq}\,\propto
\left[\mathcal{D}Nv_0^2-\sum_{j\,=1}^n\bm{v}_j^2- \frac
{\left(\sum_{j\,=1}^n\bm{v}_j\right)^2}{N-n}\right]^{(N-\,n-1)
\mathcal{D}/2-\,1} \nonumber
\end{eqnarray}
Soon we will use the first and second statistical moments of these
distributions:
\begin{eqnarray}
\langle \bm{v}_j \rangle\,=\,0\,\,, \,\,\,\,\langle v_{i\alpha}
v_{j\beta}
\rangle\,=\,v_0^2\,\delta_{\alpha\beta}\left[\delta_{ij}\,- \,\frac
{1-\delta_{ij}}{N-1}\right]\,\,,\label{vmm}
\end{eqnarray}
where $\,\langle \,[...]\,\rangle\,=\,\int...\int \,[...]\,
\,\overline{F}_n^{\,eq}\,d\bm{v}_1... \,d\bm{v}_n\,$ and $\,i,j\leq
n\,$, and besides the incomplete integral
\begin{eqnarray}
\int
\bm{v}_{n+1}\,\overline{F}_{n+1}^{\,eq}\,d\bm{v}_{n+1}\,=\,-\frac
{\bm{v}_1+...\,+\bm{v}_n}{N-n}
\,\,\overline{F}_n^{\,eq}\,\,\label{cm}
\end{eqnarray}
Obviously, factor $\,-(\bm{v}_1+...\,+\bm{v}_n)/(N-n)\,$ here is
nothing but conditional average value of $\,\bm{v}_{n+1}\,$ under
fixed $\,\bm{v}_{j\leq n}\,$.

{\bf 4.\,} As in \cite{p1}, mostly we are interested in spatial
distributions and conjugated probability flows
\begin{eqnarray}
W_n(t,\bm{R})\,=\int...\int
\overline{F}_n(t,\bm{R},\bm{v}_1...\bm{v}_n)\,d\bm{v}_1...
\,d\bm{v}_n\,\,\,,\nonumber\\
\bm{J}_n(t,\bm{R})\,=\int...\int\bm{V}^{(n)}\,
\overline{F}_n(t,\bm{R},\bm{v}_1...\bm{v}_n)\,d\bm{v}_1...
\,d\bm{v}_n\,\,\,,\nonumber
\end{eqnarray}
first of all, $\,W_1(t,\bm{R})\,$ which is probability density of
random Brownian displacement of initially localized particle (from
its start position near $\,\bm{R}=0\,$).

Similar to \cite{p1}, let us derive from (\ref{aeq}) roughened but
closed and solvable equations for these distributions, by using the
fact that after a few collisions (counted off from $\,t=0\,$)
cross-correlation between previously accumulated displacement
$\,\bm{R}\,$ of Brownian particle and its current velocity
$\,\bm{v}\,$ becomes small. Indeed, the conditional average value of
$\,\bm{v}_{\alpha}\,$ at given $\,\bm{R}\,$ is $\,\approx
R_{\alpha}/t\,$. Since $\,|\bm{R}_{\alpha}|\sim
\sqrt{2Dt}=\sqrt{2v_0^2\tau_f t}\,$ (with $\,D\,$ being diffusivity
and $\,\tau_f \sim \lambda/v_0\,$ mean free flight time), modulus of
this average is $\,\sim  v_0\sqrt{2\tau_f /t}\,$, hence already at
$\,t\sim 20\tau_f\,$ it is small enough in comparison with magnitude
$\,v_0\,$ of equilibrium velocity fluctuations.

At larger time the displacement-velocity cross correlation can be
treated as a weak perturbation. Clearly, corresponding weakly
perturbed distribution functions $\,\overline{F}_n\,$ can be written
as
\begin{eqnarray}
\overline{F}_n(t,\bm{R},\bm{v}_1...\bm{v}_n)\,=\,
\overline{F}_n^{\,eq}(\bm{v}_1...\,\bm{v}_n)\,\times\,
\,\,\,\,\,\,\,\,\,\,\,\,\label{pdf}\\
\times\,\,[\,W_n(t,\bm{R})\,+\,\bm{w}_n(t,\bm{R})\cdot
(\bm{v}_1+...\,+\bm{v}_n)\,+...\,]\,\,\,,\nonumber
\end{eqnarray}
where first term in squire brackets represents zero-order
contribution from asymptotic with independent velocity and
displacement, while the dots replace higher-order contributions. The
latter will be neglected and omitted. Substituting (\ref{pdf}) into
definition of the flows and using (\ref{vmm}) we find relations of
$\,\bm{w}_n\,$ to them:
\begin{eqnarray}
\bm{J}_n(t,\bm{R})\,=\,\frac
{N-n}{N-1}\,v_0^2\,\bm{w}_n(t,\bm{R})\,\,\label{jw}
\end{eqnarray}

Next notice that the question about connections between distribution
functions $\,\overline{F}_{n+1}^{\,in}\,$ and
$\,\overline{F}_{n+1}\,$ will be simply resolved in the framework of
the first-order approximation if we write
\begin{eqnarray}
\overline{F}_{n+1}^{\,in}(t,\bm{R},\bm{v}_1...\,\bm{v}_{n+1})\,=\,
\overline{F}_{n+1}^{\,eq}(\bm{v}_1...\,\bm{v}_{n+1})\,\times\,
\,\,\,\,\,\,\,\,\label{finc}\\
\times\,\{\,W_{n+1}(t,\bm{R})+\bm{w}_{n+1}(t,\bm{R})\cdot
a_n\,[\sum_{j\,=1}^n\bm{v}_j+c_n\bm{v}_{n+1}]\}\nonumber
\end{eqnarray}
and choose coefficient $\,a_n\,$ and $\,c_n\,$ to be such that mean
value of $\,\bm{v}_{n+1}\,$ turns into zero,
\begin{eqnarray}
\int\bm{v}_{n+1}\,\overline{F}_{n+1}^{\,in}
\,d\bm{v}_1...\,d\bm{v}_{n+1}=\,0\,\,\,, \label{win}
\end{eqnarray}
while mean values of $\,\bm{v}_j\,$ with $\,j<n+1\,$ remain exactly
the same as under distribution $\,\overline{F}_{n+1}\,$:
\begin{eqnarray}
\int\bm{v}_j\,\overline{F}_{n+1}^{\,in}
\,d\bm{v}_1...\,d\bm{v}_{n+1}= \int\bm{v}_j\,\overline{F}_{n+1}
\,d\bm{v}_1...\,d\bm{v}_{n+1}\,\nonumber
\end{eqnarray}
With the help of (\ref{vmm})-(\ref{cm}), these requirements yield
\begin{eqnarray}
c_n\,=\,\frac {n}{N-1}\,\,\,,\,\,\,\,\,a_n\,=\,\frac
{N-1}{N}\,\,\,,\label{acn}
\end{eqnarray}
respectively.

At such $\,a_n\,$ and $\,c_n\,$, expressions (\ref{finc}) and
(\ref{win}) say that velocity of outer $\,(n+1)$-th particle is not
correlated with position of $\,n$-particle cluster it came in (as if
it was a particle from equilibrium thermostat), although its position
can be correlated (as if it was full member of greater
$\,(n+1)$-particle cluster). This is just what is required by the
``weakened molecular chaos'' suggested in \cite{i1,i2} as a
compromise of mutually contradicting ``pure Stosshalansatz'' and
conservation of probabilities.

Consider time derivatives of the probability flows, combining
(\ref{aeq}) with expressions (\ref{pdf}) and (\ref{finc}), noticing
that
\begin{eqnarray}
\langle \,V^{(n)}_{\alpha}V^{(n)}_{\beta}\rangle\,=\,
\delta_{\alpha\beta}\,\frac {N-n}{n\,(N-1)}\,v_0^2\,\label{vn}
\end{eqnarray}
due to (\ref{vmm}), and everywhere keeping only lowest-order nonzero
contributions:
\begin{eqnarray}
\frac {\partial \bm{J}_n}{\partial t}\,=\,-\,\frac
{N-n}{n\,(N-1)}\,v_0^2\,
\frac {\partial W_n}{\partial \bm{R}}\,+
\,\,\,\,\,\,\,\,\,\,\,\,\,\,\,\,\,\,\,\,\,\,\,\,\nonumber\\
+\,\frac {N-n}{N}\,a_n\int d\bm{v}_1...\int
d\bm{v}_n\,\bm{V}^{(n)}\times
\,\,\,\,\,\,\,\,\,\,\,\,\label{je}\\
\,\,\,\,\times\,\,\bm{w}_{n+1}(t,\bm{R})\cdot\sum_{i\,=1}^n
\,S_{\,i\,n+1}\,[\,\sum_{j\,=1}^n\bm{v}_j+c_n\bm{v}_{n+1}]\,
\overline{F}_{n+1}^{\,eq}\,\nonumber
\end{eqnarray}

Action of the collision operator is defined as usually,
\begin{eqnarray}
S_{\,i\,n+1}\,f(...\,\bm{v}_i...\,\bm{v}_{n+1})\,=\,\nu\int
db\int d\bm{v}_{n+1}\,\times\,\,\,\,\,\,\,\,\nonumber\\
\times\,\,|\bm{v}_i-\bm{v}_{n+1}|\,[f(...\,\bm{v}_i^{\prime}...
\,\bm{v}_{n+1}^{\prime})- f(...\,\bm{v}_i...\,\bm{v}_{n+1})]\,
\,\,\,,\nonumber
\end{eqnarray}
where $\,b\,$ is $\,(\mathcal{D}-1)$-dimensional impact parameter,
and
$\,\bm{v}_i^{\prime}=\,\bm{v}_i^{\prime}(\,\bm{v}_i\,,\bm{v}_{n+1},b)\,$
and $\,\bm{v}_{n+1}^{\prime}=
\,\bm{v}_{n+1}^{\prime}(\,\bm{v}_i\,,\bm{v}_{n+1},b)\,$ are those
in-state velocities which result in given out-state velocities
$\,\bm{v}_i\,$ and $\,\bm{v}_{n+1}\,$ after collision. Taking into
account that
$\,\bm{v}_i+\bm{v}_{n+1}=\bm{v}_i^{\prime}+\bm{v}_{n+1}^{\prime}\,$
and $\,\overline{F}_{n+1}^{\,eq}\,$ are invariants of collision, it
is easy to verify that
\begin{eqnarray}
S_{\,i\,n+1}\,[\,\sum_{j\,=1}^n\bm{v}_j+c_n\bm{v}_{n+1}]\,
\overline{F}_{n+1}^{\,eq}\,=
-\frac 12\,(1-c_n)\,\nu\times\,\,\nonumber\\
\times\,\int db\int d\bm{v}_{n+1}
[\,1-\cos\,\theta\,]|\bm{v}_i-\bm{v}_{n+1}|(\bm{v}_i-\bm{v}_{n+1})
\overline{F}_{n+1}^{\,eq}
\nonumber
\end{eqnarray}
where $\,\theta =\theta(|\bm{v}_i-\bm{v}_{n+1}|\,,b)\,$ denotes
scattering angle as a function of collision parameters. Of course,
spherical symmetry of interaction potential was assumed. Now perform
in (\ref{je}) integrations first over all $\,\bm{v}_j\,$ with
$\,j\neq i,n+1\,$, using relations similar to (\ref{cm}), and second
over $\,\bm{v}_i\,$ and $\,\bm{v}_{n+1}\,$, paying attention to
symmetry of resulting equations, and introduce the ``relaxation
rate''
\begin{eqnarray}\gamma\,=\,\frac
{(N-1)\,\nu}{4Nv_0^2}\int d\bm{v}\int d\bm{v}^{\prime}\int
db\,\,|\bm{v}-\bm{v}^{\prime}|^3\,\times\nonumber\\
\times\,[\,1-\cos\,\theta (|\bm{v}-\bm{v}^{\prime}|\,,b)\,]\,
\overline{F}_{2}^{\,eq}(\bm{v},\bm{v}^{\prime})\,\nonumber
\end{eqnarray}

Finally, with the help of (\ref{jw}) and (\ref{acn}), equation
(\ref{je}) transforms to
\begin{eqnarray}
\frac {\partial \bm{J}_n}{\partial t}\,=\,-\,\frac {v_0^2}{n}\,\frac
{N-n}{N-1}\, \frac {\partial W_n}{\partial \bm{R}}\,-\,\gamma\,\frac
{N-n}{N}\,\,\bm{J}_{n+1}\,\,\,,\label{fe}
\end{eqnarray}
It remains to supplement (\ref{fe}) by equations
\begin{eqnarray}
\frac {\partial W_n}{\partial t}\,=\,-\,\frac {\partial
\bm{J}_n}{\partial \bm{R}}\,\,\,,\label{pe}
\end{eqnarray}
and initial conditions
\begin{eqnarray}
W_n(t=0,\bm{R})\,=\,n\,W(\bm{R})\,\,\,,
\,\,\,\bm{J}_n(t=0,\bm{R})=0\,\,\,, \label{ic}
\end{eqnarray}
which directly follow from (\ref{aeq}) and (\ref{aind}).

Clearly, $\,\gamma\,$ can be identified with inverse mean free flight
time $\,1/\tau_f\,$. We may replace $\,W(\bm{R})\,$ by delta-function
$\,\delta^{\mathcal{D}}(\bm{R})\,$, since anyway we have to consider
spatial scales much greater then $\,d\,$. Of course, index $\,n\,$ in
(\ref{fe})-(\ref{ic}) takes values $\,1\leq n\leq N-1\,$, because
zero total momentum in our micro-canonical statistical ensemble
(\ref{ind}), (\ref{aind}) requires to set
\begin{eqnarray}
\bm{J}_N(t,\bm{R})=0\,\,\, \label{jN}
\end{eqnarray}
The same is dictated by (\ref{jw}), (\ref{vn}) or (\ref{fe}) at
$\,n=N\,$.

If we chose canonical ensemble, then in (\ref{win}) we would have to
write $\,c_n=0\,$ and $\,a_n=1\,$ in place of (\ref{acn}), while
formulas (\ref{jw}) and (\ref{vn}) would change to $\,\bm{J}_n
=v_0^2\bm{w}_n\,$ and $\,\langle
V^{(n)}_{\alpha}V^{(n)}_{\beta}\rangle =
\delta_{\alpha\beta}\,v_0^2/n\,$, resulting in
\begin{eqnarray}
\frac {\partial \bm{J}_n}{\partial t}\,=\,-\,\frac {v_0^2}{n}\,\frac
{\partial W_n}{\partial \bm{R}}\,-\,\gamma\,\frac
{N-n}{N}\,\,\bm{J}_{n+1}\,\,\,,\label{fec}
\end{eqnarray}
instead of (\ref{fe}), now with $\,1\leq n\leq N\,$ and
$\,\bm{J}_N\neq 0\,$.

\section{Model analysis}

{\bf 1.\,} First, as in \cite{p1}, consider solution to equations
(\ref{fe})-(\ref{ic}) in terms of Fourier and Laplace transforms
\begin{eqnarray}
\Xi_n(t,\,\bm{k})\,=\,\int\exp(i\,\bm{k}\cdot
\bm{r})\,W_n(t,\bm{r})\,d\bm{r}\,\,\,,\nonumber\\
\Phi_n(p,\,\bm{k})\,=\,\int_0^\infty
\exp(-p\,t)\,\,\Xi_n(t,\,\bm{k})\,dt\,\,\nonumber
\end{eqnarray}
At that, in view of obvious isotropy of Brownian motion under
interest, it is sufficient to watch for its projection onto some axis
only, formally changing ``wave vector'' $\,\bm{k}\,$ by scalar
$\,k\,$. Then
\begin{eqnarray}
\Phi_1(p,k)=\,\frac 1p\,+\sum_{n=1}^\infty \frac
{(-k^2)^n}{(2n)!}\int_0^\infty e^{-p\,t}\,\langle R^{2n}(t)\rangle
\,dt\,\nonumber
\end{eqnarray}
with $\,\langle R^{2n}(t)\rangle =\int R^{2n}\,W_1(t,R)\,dR\,$ being
statistical moments of (a projection of) Brownian displacement.

Now from (\ref{fe})-(\ref{ic}) we come to finite series:
\begin{eqnarray}
p\,\Phi_n(p,\,k)\,-\,n\,=\,\frac {v_0^2k^2 N}{p\,\gamma
(N-1)}\,\times \label{sn}\,\,\,
\,\,\,\,\,\,\,\,\,\,\,\,\,\,\,\,\,\,\,\,\,\,\,\,\\
\times \,\sum_{s\,=n}^{N-1} \prod_{m\,=n}^s \left[-\frac {(N-m)\gamma
}{Np} \right]\left[1+\frac {v_0^2k^2 (N-m)}{m
p^{\,2}(N-1)}\right]^{-1}\,\nonumber
\end{eqnarray}
Correspondingly, Laplace transform of $\,\langle R^{2n}(t)\rangle\,$
is $\,(2n+N-1)$-order polynomial of $\,1/p\,$ with powers from
$\,2n+1\,$ to $\,2n+N-1\,$. Hence, $\,\langle R^{2n}(t)\rangle\,$
itself is $\,(2n+N-2)$-order polynomial of $\,t\,$ with powers from
$\,2n\,$ to $\,2n+N-2\,$.

In particular,
\begin{eqnarray}
\int_0^\infty e^{-p\,t}\langle R^{2}(t)\rangle \,dt=-\frac
{2D}{p^{\,2}}\sum_{s=1}^{N-1}\frac {(N-1)!}{(N-s-1)!}\left(-\frac
{\gamma}{Np}\right)^s\nonumber
\end{eqnarray}
which means that
\begin{eqnarray}
\frac {d}{dt}\,\langle R^2(t)\rangle =\,2D\left[\,1\,-
\,\left(1-\frac {\gamma t}{N}\right)^{N-1}\right]\,\label{r2}
\end{eqnarray}
with diffusivity $\,D\,$ presented by
\begin{eqnarray}
D\,=\,\frac {v_0^2N}{\gamma (N-1)}\,\,\nonumber
\end{eqnarray}

Consider also fourth-order moment of (projection of) the
displacement:
\begin{eqnarray}
\int_0^\infty e^{-p\,t}\langle R^{4}(t)\rangle
\,dt\,=\,-\,\frac {24D^2\gamma}{p^{\,4}}\,\times
\,\,\,\,\,\,\,\,\,\,\,\,\,\nonumber\\
\times\,\sum_{s=1}^{N-1}\frac {(N-1)!}{(N-s-1)!}\left(-\frac
{\gamma}{Np}\right)^s\left[\,\sum_{m=1}^s\frac 1m\,-\frac
sN\right]\nonumber
\end{eqnarray}
which yields in the time domain
\begin{eqnarray}
\frac {d^{\,2}}{dt^{2}}\,\langle R^4(t)\rangle
=\,24D^2\left(\,\vartheta^{N-1}-\frac 1N -\frac
{N-1}{N}\,\vartheta^{N}\right)\,+\nonumber\\
+\,\,24D^2\sum_{s\,=1}^{N-1}\left(\frac {1-\vartheta^{\,s}}{s}-\frac
{\vartheta^{\,s}-\vartheta^{N}}{N-s}\right)\,\,\,\,\,\,\,\,\,\label{r4}
\end{eqnarray}
with factor $\,\vartheta =\vartheta (t)\,$ defined by
\begin{eqnarray}
\vartheta\,=\,1-\frac {\gamma t}{N}\nonumber
\end{eqnarray}

Polynomials representing higher-order statistical moments may be
considered in terms of functions $\,V^{(m)}_n(t)\,$ introduced by
analogy with right-hand sides of (\ref{r2})-(\ref{r4}):
\begin{eqnarray}
2m\,\left(\frac {d\,}{dt}\right)^{m-1} \int
R^{2m-1}\,J_{\,n}(t,R)\,dR\,=\,\,\,\,\,\,\,\,\,\,\,
\,\,\,\,\,\,\,\label{vs}\\
=\,\left(\frac {d\,}{dt}\right)^m \int
R^{2m}\,W_n(t,R)\,dR\,\equiv\,(2m)!\,D^m\,\,V^{(m)}_n(t)\,
\,\nonumber
\end{eqnarray}
From (\ref{fe})-(\ref{ic}) recurrent equations for $\,V^{(m)}_n(t)\,$
directly follow:
\begin{eqnarray}
\frac {d V^{(m)}_{\,n}}{d t}\,=\,\gamma\, \frac {N-n}{N}
\left[\,\frac {1}{n}\,\,V^{(m-\,1)}_n\,-\,V^{(m)}_{n+1}
\,\right]\label{vm}
\end{eqnarray}
with $\,1\leq n\leq N-1\,$, $\,V^{(0)}_n(t)\,=\,n\,$, $\,m\geq1\,$,
initial conditions $\,V^{(m)}_n(t=0)=0\,$, and $\,V^{(m)}_N=0\,$
because of (\ref{jN}). In particular, at $\,m=1\,$ equations
(\ref{vm}) yield generalization of (\ref{r2}),
\begin{eqnarray}
\frac {d}{dt}\int
R^2\,W^{(1)}_n\,dR\,=\,2D\,\left[1-\vartheta^{N-\,n}\right]\,
\label{r2n}
\end{eqnarray}

{\bf 2.\,} Expressions (\ref{r2}) and (\ref{r4}) look quite formally
satisfactory and physically reasonable as long as $\,\gamma t<N\,$,
that is $\,\vartheta >0\,$. Moreover, as it naturally might be
expected, at sufficiently large $\,N\,$ and $\,1\ll \gamma t<N\,$
(more precisely, at $\,|\vartheta |^N< 1\,$) they approximately
coincide with expressions for infinite system \cite{p1}, that is
$\,2D\,$ and $\,24D^2\,\ln\,\gamma t\,\,$, respectively (as the
consequence, while $\,\gamma t< N\,$ the diffusivity $\,1/f$-noise
exists in the sense explained in \cite{p1,i1}). Similar statements
are true also in respect to various higher moments $\,\int
R^{2m}W_n(t,R)\,dR\,$ which follow from (\ref{sn}) or (\ref{vm}) and
to the whole probability distribution functions $\,W_n(t,R)\,$ (at
least at $\,N-n\gg 1\,$). Particular example is shown in Fig.1.

\begin{figure}
\includegraphics{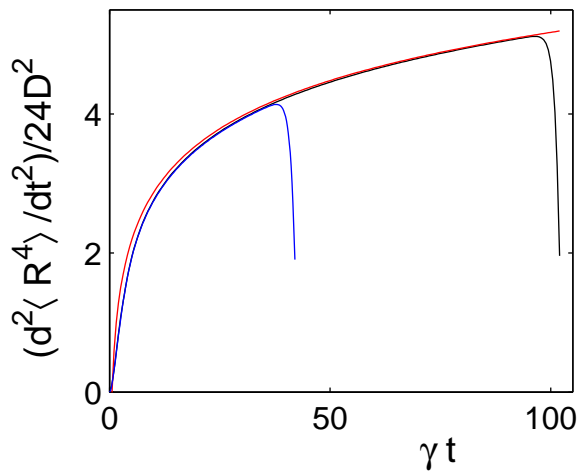}
\caption{\label{fig1} Plots of $\,(24D^2)^{-1}\,d^2\langle
R^4(t)\rangle/dt^2\,$ via $\,\gamma t\,$ corresponding to (\ref{r4})
at $\,N=21\,$ (blue) and $\,N=51\,$ (black) in comparison with
$\,\ln{(\gamma t)}+C\,$ (red) which approximates limit
$\,N\rightarrow \infty\,$ ($\,C\,$ is Euler constant).}
\end{figure}

However, at $\,\gamma t>N\,$ ($\,\vartheta <0\,$) the same
expressions (\ref{r2}), (\ref{r4}) and (\ref{r2n}) become obviously
senseless. Notice that $\,\vartheta \,$ and its various powers
$\,\vartheta ^{\,s}\,$ play roles of relaxation factors being direct
analogues,  especially  at $\,s\sim N\,$, of the exponent
$\,\exp{(-\gamma t)}\,$ in infinite system. Therefore, negative
$\,\vartheta \,$ with growing absolute value has none physical
meaning.

Hence, we should conclude that solution to equations of our model
(\ref{fe})-(\ref{ic}) possesses peculiar time point $\,N/\gamma
=N\tau_f\,$. Out of it the model does not work.

These fact can not be addressed to some incidental details of the
model. Indeed, other model corresponding to canonical ensemble and
represented by equations (\ref{fec}) in place of (\ref{fe}) possesses
the same peculiar point. It is not hard to trace that the peculiarity
arises from two things: presence of factor $\,N-n\,\,$ in collision
term of equations (\ref{fe}) and characteristic open-chain structure
of these equations with $\,n$-th member referring to $\,(n+1)$-th.
Both have migrated from the basic BBGKY equations. But the second is
also child of the ``weakened molecular chaos'' hypothesis. Hence,
just the latter (to be precise, its concrete realization on the form
of (\ref{finc})-(\ref{win})) is responsible for the peculiarity.

It seems natural to interpret this situation as evidence that at
$\,\gamma t= N\,$ period of non-Gaussian behavior of our random walk
is over. Then predicted peculiar time $\,N/\gamma \,$ conceptually as
well as numerically coincides with time $\,\tau_{ng} \,$ suggested in
\cite{p1} as characteristic time scale separating ``non-ergodic''
(non-Gaussian) and ``ergodic'' (asymptotically Gaussian) stages of
the walk.

{\bf 3.\,} In order to construct a better model (where weakened
molecular chaos would self-destruct with time into complete one) we
should return to higher level of equations (\ref{aeq}) or BBGKY
equations. This is subject of separate investigation (principally, it
is plausible if ``trivial'' Gaussian asymptotic creates nontrivial
problems).

At present, we confine ourselves by discussion of most weak spot of
the model. Undoubtedly, that is the ``end cap'' (\ref{jN}) which
makes closure of the chain (\ref{fe}), but rather bad closure,
because it means that $\,(N-1)$-particle cluster loses any
interaction with rest $\,N$-th particle. Here is the reason to remove
$\,(N-1)$-th of equations (\ref{fe}) from affected zone of the
``weakened molecular chaos'' hypothesis (\ref{win})-(\ref{acn}). Then
it takes the form
\begin{eqnarray}
\frac {\partial \bm{J}_{N-1}}{\partial t}=-\,\frac {v_0^2}{(N-1)^2}\,
\frac {\partial W_{N-1}}{\partial \bm{R}}\,- \,\gamma\,\frac
{1}{N}\,\widetilde{\bm{J}}_{N}\,\,\,,\nonumber\\
\widetilde{\bm{J}}_{N}\,\equiv\,
v_0^2\,a_{N-1}(1-c_{N-1})\,\bm{w}_{N}\,\,\,\,\,\,\,\,\,\,\,\,\,\,\,\,
\label{fen}
\end{eqnarray}
Equality (\ref{jN}) remains true, but now it is out of work. The
model becomes temporarily non-closed, and we get chance to correct
its behavior after $\,t\sim \tau_{ng}\,$ (notice that $\,\tau_{ng}\,$
is just time necessary to transfer influence by
$\,\widetilde{\bm{J}}_{N}\,$ from last to first of equations
(\ref{fe})).

Let us demonstrate that in presence of $\,\widetilde{\bm{J}}_{N}\,$
the model comprises Gaussian asymptotic at
$\,t/\tau_{ng}\rightarrow\infty\,$, moreover, without significant
assumptions about accompanying asymptotic of
$\,\widetilde{\bm{J}}_{N}\,$. According to (\ref{vs}), Gaussian
asymptotic means that $\,V^{(m)}_1(t\rightarrow\infty)\rightarrow
1\,$. Then other variables in equations (\ref{vm}) also have fixed
points fully determined by $\,V^{(m)}_1(\infty)\,$ (as combined with
$\,V^{(0)}_n=n\,$):
\begin{equation}
V^{(m)}_{n}(\infty )\,=\,\left \{\nonumber%
\begin{array}{ll}
    (n-m)!/(n-1)! & \hbox{,} \,\,\,\,1\leq m< n\,\,\, \\ \nonumber
    V_1^{(m-\,n\,+1)}(\infty)/(n-1)! & \hbox{,} \,
    \,\,\,m\geq n\,\,\, \nonumber
\end{array}%
\right.    \,\,\label{ga}
\end{equation}
Here $\,1\leq n\leq N\,$, and quantities $\,V_N^{(m)}(t)\,$ represent
$\,\widetilde{J}_{N}(t,R)\,$ by the common rule (\ref{vs}) with
$\,n=N\,$ and $\,\widetilde{J}_{N}(t,R)\,$ in place of
$\,J_{N}(t,R)\,$. We did not fulfil substitution $V^{(m)}_1(\infty)=
1$. This helps to visualize that $V^{(m)}_N(\infty)$ at least at
$m<N$ appear insensible to $V^{(m)}_1(\infty)$ (i.e. independent on
assumption about gaussianity of the asymptotic) thus justifying above
statement. Notice also that equalities $V^{(m)}_1(\infty)= 1$ imply
equalities $\,V^{(m)}_2(\infty)=V^{(m)}_1(\infty)\,$ (at $m>0$),
which mean that asymptotically $\,J_2(t,R)\rightarrow J_1(t,R)\,$ and
therefore first pairs of equations (\ref{fe})-(\ref{pe}) stands
apart, as it must be in standard kinetics.

\section{Conclusion}
In the first part of this paper \cite{p1} a simple kinetic model of
self-diffusion (Brownian motion) in infinite-volume gas was
developed. It exploits collisional description of interaction between
particles but declines Boltzmann's ``molecular chaos'' hypothesis
because the latter appears incompatible with two doubtless
theoretical requirements to be fulfilled: probability conservation
during collisions and necessity to deal with spatially nonuniform
statistical ensemble when considering random walks of gas particles.
Instead, the ``weakened molecular chaos'' hypothesis suggested in
\cite{i1} was used, which is consistent with both the requirements
and claims statistical independency of colliding particles in respect
to their velocities only but not their coordinates. It resulted in a
chain of kinetic equations whose solution shows that statistics of
random walks of gas particles does not obey the law of large numbers
and remains essentially non-Gaussian at arbitrary long time scales,
including what can be named 1/f fluctuations in diffusivity.

Physical origin of so wild statistical freedom is absence of {\it
actual} cause-and-effect correlations between successive fragments of
the random walk, in spite of {\it statistical} long-living
correlations formally describing the same freedom \cite{i1,i2}. Its
mathematical origin is absence of relaxation (in the ordinary sense)
terms in kinetic equations. Usual place of relaxation terms is
occupied by references to next higher-order equations. This chain of
references mean that a particle whose random walk is under
observation constantly accumulates (actual) correlations with more
and more new particles.

In infinite gas this process lasts endlessly. In finite
$\,N$-particle gas it ends at time $\sim N$ (counted from start of
the observation, in units of mean free flight time) when collisions
of observed particle with all others have realized, therefore,
initial conditions of the whole system have snapped into action.
Further motion of this particle is fully predestined by already
observed walk, representing its reflections (mappings) in complete
non-decaying actual correlation with all the gas. This creates the
necessary prerequisites for decay of statistical correlations between
particles and realization of the law of large numbers and Gaussian
statistics at next longer time scales.

The model investigated in this second part of the paper certainly
detects time $\approx N$ of this transition but, unfortunately, is
unfit for description of the next Gaussian asymptotic of the random
walk. However, in principle, the exact BBGKY equations give a firm
base for proper improving the model. From the other hand, practically
sooner just the former non-Gaussian stage is of interest, since $N$
means rather large time. In this respect, the model needs in 
improvements too. Probably, at present form it somehow overestimates
``degree of non-gaussianity'' of the random walk. The early
phenomenological construction of ``quasi-Gaussian'' walk \cite{bk3}
(see also references therein and \cite{i2}) produced much softer
non-gaussianity (although also with 1/f noise in diffusivity). In any
case, there are many interesting questions for future.



\begin{thebibliography}{6}

\bibitem{p1}
Yu.\,E.\,Kuzovlev, ``On statistics and 1/f noise of Brownian motion
in Boltzmann-Grad gas and finite gas on torus. II. Infinite gas'',
arXiv: cond-mat/0609515.

\bibitem{i1}
Yu.\,E.\,Kuzovlev, ``BBGKY equations, self-diffusion and 1/f-noise in
a slightly nonideal gas'', Sov.Phys.-JETP, {\bf 67} (12), 2469 (1988)
[in Russian: Zh.Eksp.Teor.Fiz., {\bf 94}, No.12, 140 (1988)].

\bibitem{i2}
Yuriy\,E.\,Kuzovlev, ``Kinetical theory beyond conventional
approximations and 1/f-noise'', arXiv: cond-mat/9903350.

\bibitem{kac}
M.\,Kac, ``Probability and related topics in physical sciences'',
Intersci. Publ., London, N.-Y., 1957.

\bibitem{rdl}
P.\,Resibois and M.\,de\,Leener. Classical kinetic theory of fluids.
Wiley, New-York, 1977.

\bibitem{bk3}
G.\,N.\,Bochkov and Yu.\,E.\,Kuzovlev, ``New in 1/f-noise studies'',
Sov.Phys.-Usp., {\bf 26}, 829 (1983) [in Russian: UFN, {\bf 141}, 151
(1983)].

\end{thebibliography}


\end{document}